\documentclass[12pt,a4paper,english]{article}
\usepackage{amsmath}
\usepackage{amssymb}
\usepackage{amsthm}

\usepackage[authoryear]{natbib}
\usepackage{booktabs}
\usepackage{graphicx}
\usepackage{epstopdf}
\usepackage{caption}
\usepackage{hyperref}
\usepackage{titling}
\usepackage{xcolor}
\usepackage{babel}
\usepackage[title]{appendix}
\usepackage[onehalfspacing]{setspace}

\definecolor{lightblue}{rgb}{0, 0.4, 0.6}
\hypersetup{
    breaklinks=true,
    colorlinks = true,
    citecolor = {lightblue},
    linkcolor = {lightblue},
}
\usepackage[capitalise,nameinlink
,noabbrev]{cleveref}
\crefname{equation}{}{}
\crefname{enumi}{}{}
\crefname{appsec}{Appendix}{Appendices}
\crefname{appsubsec}{Appendix}{Appendices}
\crefname{algocf}{Algorithm}{Algorithms}
\crefname{table}{Table}{Tables}
\crefname{assumption}{Assumption}{Assumptions}

\usepackage[linesnumbered,ruled,vlined]{algorithm2e}
\usepackage{subcaption}
\usepackage[inline]{enumitem}

\newcommand{\pto}{\stackrel{p}{\longrightarrow}}

\newcommand{\bs}[1]{\boldsymbol{#1}}

\theoremstyle{remark}
\newtheorem{nondecr}{Theorem}
\newtheorem{theorem}[nondecr]{Theorem}
\newtheorem{nondecrl}{Lemma}
\newtheorem{lemma}[nondecrl]{Lemma}

\newtheorem{nondecra}{Assumption}
\newtheorem{assumption}[nondecra]{Assumption}
\usepackage{sectsty}
\subsubsectionfont{\normalfont\itshape}
\subsectionfont{\normalfont\itshape\bfseries}

\newcounter{bean}

\title{\vspace{-2cm}\textbf{Uniform Inference in Linear Error-in-Variables Models: Divide-and-Conquer}}
\author{}
\date{\today}
\author{Tom Boot\thanks{Corresponding author. University of Groningen, \textsf{t.boot@rug.nl}} \and Art\={u}ras Juodis\thanks{University of Amsterdam and Tinbergen Institute, \textsf{a.juodis@uva.nl}.}}

\begin{document}

\maketitle
\begin{abstract}
  \noindent It is customary to estimate error-in-variables models using higher-order moments of observables. This moments-based estimator is consistent only when the coefficient of the latent regressor is assumed to be non-zero. We develop a new estimator based on the divide-and-conquer principle that is consistent for any value of the coefficient of the latent regressor. In an application on the relation between investment, (mismeasured) Tobin's $q$ and cash flow, we find time periods in which the effect of Tobin's $q$ is not statistically different from zero. The implausibly large higher-order moment estimates in these periods disappear when using the proposed estimator.
      \textit{Word count}: 7349.
\end{abstract}

\newpage
 \section{Introduction}

    To account for measurement error in independent variables, higher-order moment estimators have been used to analyse the relation between R\&D expenditures and patent applications (\citealp{lewbel1997constructing}), to test the $q$-theory of investment in finance (\citealp{erickson2000measurement}), and to investigate firm saving behaviour (\citealp{riddick2009corporate}). Underlying these estimators are two crucial assumptions: first, there needs to be sufficient skewness in the latent regressor. Second, the coefficient $\beta_{0}$ that relates the latent regressor to the observed outcome cannot equal zero. In this paper we focus on the situation where this second assumption potentially fails.
    In particular, we consider the standard third-order moment estimator that can be attributed to \citet{geary1941inherent}. When $\beta_{0}=0$, this estimator converges to a ratio of correlated (mean zero) normal random variables.

    As a solution to this problem, we propose an estimator based on a divide-and-conquer strategy: the data is split into equal sized blocks. In each block, we calculate the denominator and numerator of the estimator (by \citealp{geary1941inherent}) on non-overlapping subsets of the data. This breaks the dependency between the numerator and denominator, and ensures that the estimator is median unbiased regardless of the value of $\beta_{0}$. Subsequently taking the median over the estimators from the different blocks yields a consistent and asymptotically normal estimator irrespective of the value of $\beta_{0}$. However, the rate of convergence of the proposed estimator depends on the number of blocks when $\beta_{0}=0$ and increases to the standard rate when $\beta_{0} \neq 0$.

    To test the proposed estimator in practice, we revisit an empirical test of corporate finance's $q$-theory. This theory states that investment fluctuations are driven by the marginal $q$: the market value of capital relative to the shadow value of capital.\footnote{An overview of the history of $q$-theory can be found in \citet{erickson2000measurement}} Empirically, $q$-theory appeared discredited with for example \citet{blundell1992investment} finding a significant role for internals funds after controlling for $q$. However, \citet{erickson2000measurement} show that these findings can be explained by the substantial measurement error in Tobin's $q$ measure of marginal $q$. This measurement error drives down the estimated coefficient on $q$, while increasing the coefficient of controls such as internal funds. Accounting for measurement error via the use of higher-order moment estimators showed that $q$-theory is not at odds with the data, see also \citet{erickson2012treating} and \citet{andrei2019did}.

    We first consider the simulation set-up of \citet{erickson2014minimum}, which is geared to the environment in which $q$-theory can be tested. We find that the divide-and-conquer estimator performs well regardless of the value of $\beta_{0}$, unlike the third-order moment estimator that is ill-defined when $\beta_{0}=0$.

    We then analyse the data on firm investment, Tobin's $q$ and cash flow from \citet{erickson2014minimum}. When using the full sample of available data, ranging from 1970 to 2011, the divide-and-conquer estimates are in line with those in \citet{erickson2014minimum}. In particular, we do not find evidence for the effect of cash flow on firm investment. Motivated by the notion of \citet{erickson2014minimum} that there exists variation over time in the estimates, we then re-estimate the coefficients using an expanding window of data starting with the data in 1970-1980 and ending with the full sample 1970-2011. We indeed find substantial changes in the estimated relation between investment, Tobin's $q$ and cash flow over time. In particular, we find rather extreme point estimates and confidence intervals in time periods before the mid-1980s. These occur precisely in time periods in which the divide-and-conquer estimator is not significantly different from zero. The divide-and-conquer estimates for the effect of Tobin's $q$ is found to be rather stable over time.

    The paper relates to three different strands of the literature. First, the error-in-variables model has been thoroughly studied in econometrics and statistics. Estimation based on third-order moments as the one we employ can be subscribed \citet{geary1941inherent}. Identification in the error-in-variables model is considered by (among others) \citet{reiersol1950identifiability,kapteyn1983identification,bekker1986comment}. \citet{pal1980consistent} extends this estimator for the single regressor case based on higher-order cumulants, while \citet{dagenais1997higher}, \citet{cragg1997using} and \citet{lewbel1997constructing} consider other functions of mismeasured regressors. An overview of the literature can be found in \citet{wansbeek2000measurement} and more recently \citet{schennach2016recent}.

    A general framework for higher-order moment estimators in models with multiple mismeasured and perfectly measured regressors is proposed in \citet{erickson2002two}. Instead of the moments, it is somewhat simpler to rely on higher-order cumulants as in \citet{erickson2014minimum} as the estimators are available in closed-form. Nonparametric identification and semiparametric estimation are considered in \citet{schennach2013nonparametric}.

    The second strand of literature concerns divide-and-conquer estimators. These estimators are developed for settings where a massive data set cannot be loaded into memory, see for instance \citet{shi2018massive}. The idea is to construct a sequence of estimators on independent subsets of the data, and then aggregate the estimators into a single estimator. This is sometimes also referred to as distributed inference. The common aggregation method is to take the mean. Applied to monotone regression, \citet{banerjee2019divide} document a superefficiency property of this method. Distributed quantile regression is studied by \citet{chen2019quantile} and \citet{volgushev2019distributed}.

    Finally, rather than taking the mean of the subsample estimators, we rely on the median. This is shared with median-of-means (MoM) estimators that are used to robustify machine learning algorithms in the presence of heavy-tailed data. MoM estimators were originally proposed by \citet{nemirovskij1983problem}, \citet{jerrum1986random}, and \citet{alon1999space}. In recent years, various authors show that these estimators attain optimal rates of convergence under weak assumptions on the data  (\citealp{hsu2014heavy}, \citealp{lugosi2019risk}, and \citealp{lecue2020robust}). However, theoretical results are only available under the assumption that the random variables have finite second moment. The estimator we propose is closer to a median-of-ratios-of-means estimators, which in the worst case scenario does not have any finite moments.

    The paper proceeds as follows. In \cref{sec:model} we describe the model, the proposed estimator and its implementation. In \cref{sec:theory} we describe asymptotic results for the divide-and-conquer estimator. The simulation study and empirical application are presented in \cref{sec:numerical}. \cref{sec:conclusion} concludes. Proofs are deferred to \cref{app:proofs}.

    \section{Model, Estimators and Inference}\label{sec:model}
    \subsection{A Simple Error-in-Variables Model}
    As a basis of our analysis we consider the following simplified error-in-variables model for one variable
    \begin{equation}\label{eq:eiv}
        \begin{split}
            y_{i} &= \xi_{i}\beta_{0} + \varepsilon_{i},\\
            x_{i} & = \xi_{i} + u_{i},
        \end{split}
    \end{equation}
    for $i=1,\ldots, n$.
    Here the observed variables are $(y_{i},x_{i})$, while the remaining variables are latent. The main parameter of interest is $\beta_{0}$, the causal effect of marginal change in $\xi_{i}$ on $y_{i}$. As $\xi_{i}$ is not observed, one could naively consider estimating $\beta$ by running the OLS regression of $y_{i}$ on $x_{i}$:
    \begin{equation}\label{eq:ols}
        \hat{\beta}^{OLS} = \frac{\frac{1}{n}\sum_{i=1}^{n}x_{i}y_{i}}{\frac{1}{n}\sum_{i=1}^{n}x_{i}^2}.
    \end{equation}
    This estimator, however, generally is inconsistent as in the limit $n\to\infty$:
    \begin{equation}
        \hat{\beta}^{OLS}\pto \beta_{0}\frac{\sigma_{\xi}^{(2)}}{\sigma_{u}^{(2)}+\sigma_{\xi}^{(2)}},
    \end{equation}
    provided that all stochastic quantities are mutually independent. Here we denote $\mathbb{E}[\xi^2_{i}] = \sigma_{\xi}^2$, $\mathbb{E}[\xi_{i}^3] = \sigma_{\xi}^{(3)}$, $\mathbb{E}[\xi_{i}^4] = \sigma_{\xi}^{(4)}$ and similarly for $\varepsilon_{i}$ and $u_{i}$ we have $\mathbb{E}[\varepsilon_{i}^2] = \sigma_{\varepsilon}^2$, $\mathbb{E}[\varepsilon_{i}^3] = \sigma_{\varepsilon}^{(3)}$, $\mathbb{E}[\varepsilon_{i}^4] = \sigma_{\varepsilon}^{(4)}$, $\mathbb{E}[u_{i}^2] = \sigma_{u}^{(2)}$.

    As the OLS estimator is generally inconsistent, in what follows we limit our attention to a class of method-of-moments estimators that use the information from the higher order moments of the data. The parameter $\beta_{0}$ can then be estimated with the estimator attributed to \citet{geary1941inherent}.

    In what follows, we assume that all random variables are mutually independent.

    \begin{assumption}\label{ass:errors}
        $(\varepsilon_{i},u_{i},\xi_{i})$ are mutually independent random variables with expectation zero.
    \end{assumption}

    A direct implication of \cref{ass:errors} is that in population
    \begin{equation}
        \label{eq:moments}
        \begin{split}
            \mathbb{E}[x_{i}y_{i}^2] &= \sigma_{\xi}^{(3)}\beta_{0}^2,\quad
            \mathbb{E}[x_{i}^2y_{i}] =\sigma_{\xi}^{(3)}\beta_{0}.
        \end{split}
    \end{equation}
    This implies the following moment condition,
    \begin{equation}
        \mathbb{E}[g_{i}(\beta)]=\mathbb{E}[x_{i}y_{i}^2 -x_{i}^2y_{i}\beta]=0,
    \end{equation}
    for $\beta=\beta_{0}$. Associated with the above population orthogonality conditions is the following estimator, attributed to \citet{geary1941inherent},
    \begin{equation}\label{eq:gmm}
        \hat{\beta}^{3M} = \frac{\frac{1}{n}\sum_{i=1}^{n}x_{i}y_{i}^2}{\frac{1}{n}\sum_{i=1}^{n}x_{i}^2 y_{i}}.
    \end{equation}
    As it is evident from \eqref{eq:moments}, the identifying power of this estimator is sensitive to the exact values of $\beta_{0}$ and/or $\sigma_{\xi}^{(3)}$. In particular, if $\beta_{0}=0$ and/or $\sigma_{\xi}^{(3)}=0$, the probability limit of the estimator is ill-defined.

    This is also evident from the expression for the asymptotic variance given by
    \begin{equation}\label{eq:asyvar}
        \begin{split}
            \mathbb{V}(\hat{\beta}^{3M};\beta) &=\frac{\mathbb{E}[g_{i}(\beta_{0})^2]}{\mathbb{E}\left[\partial g_{i}(\beta)/\partial\beta|_{\beta=\beta_{0}}\right]^2}= \frac{\mathbb{E}[x_{i}^2y_{i}^2(y_{i}-x_{i}\beta_{0})^2]}{\mathbb{E}[x_{i}^2y_{i}]^2}\\
            &=\frac{\sigma_{\xi}^{(4)}\sigma_{\varepsilon}^2\beta_{0}^2 + \sigma_{\xi}^2\sigma_{\varepsilon}^{(4)} + 2\sigma_{\xi}^{(3)}\sigma_{\varepsilon}^{(3)}\beta_{0}+ \sigma_{u}^2\sigma_{\varepsilon}^2\sigma_{\xi}^2\beta_{0}^2 + \sigma_{u}^2\sigma_{\varepsilon}^{(4)}}{[\sigma_{\xi}^{(3)}]^2\beta_{0}^2}.
        \end{split}
    \end{equation}
    Clearly, when $\beta_{0}=0$ or $\sigma_{\xi}^{(3)}=0$, the variance is ill-defined.

    \subsection{The Divide-and-Conquer Approach}
    \label{section::divide}
    In what follows, we describe the main intuition behind the procedure put forward in this paper. Notice that when $\beta_{0}=0$, the estimator \eqref{eq:gmm} is the ratio of two correlated sums that both have expectation zero. Borrowing on the idea of the Split Sample IV estimator of \citet{angrist1995split}, we can easily break the dependence between the limiting value of the numerator and the denominator by evaluating the denominator and numerator on independent samples. As a result, the resulting estimator will be centered at the true value $\beta_{0}=0$. However, the estimator will still remain non-normal in the limit for $\beta_{0}=0$, and normal for $\beta_{0}\neq 0$, invalidating any standard inference approaches.

    The above problem, however, can be solved. Define two subsets $R_{j,1}\subset \{1,\ldots,n\}$ and $R_{j,2}\subset\{1,\ldots,n\}$ such that $R_{j,1}\cap R_{j,2}=\emptyset$, denote $|R_{j,1}|=|R_{j,2}|=b/2$. The index $j$ will be used to index different choices of the subsets. Consider now the estimator,
    \begin{equation}\label{eq:subset}
        \hat{\beta}_{j,b} = \frac{\sum_{i\in R_{j,1}}{x}_{i}{y}_{i}^2}{\sum_{i\in R_{j,2}}{x}_{i}^2{y}_{i}}.
    \end{equation}
    When $\beta_{0}=0$, \eqref{eq:subset} is the ratio of two independent sums that both have expectation zero. Intuitively, we would expect that as $b\rightarrow\infty$, $\hat{\beta}_{j}$ converges weakly to a Cauchy-type random variable with the CDF $F_{c}(x)$.
    Let $F_{j,b}(x)$ be the corresponding CDF of $\hat{\beta}_{j,b}$, then in \cref{app:proofs}, we show that indeed
    \begin{equation}
        \sup_{x\in \mathbb{R}}\left|F_{j,b}(x)-F_{c}(x)\right| \leq M/\sqrt{b},
    \end{equation}
    for some constant $M>0$ when $\beta_{0}=0$. Note that as Cauchy-type random variables are symmetric, their median is $0$, which happens to be exactly the value $\beta_{0}=0$ we are after.

    In cases when the underlying data is a pure cross-section, this facilitates the following strategy to recover the median (thus also $\beta_{0})$ of the limiting random variable with the CDF $F_{c}(x)$. Randomly partition the $n$ observations into $B$ blocks, so that the number of observations within each block equals $b=n/B$. Within each block, split the sample to form \eqref{eq:subset}, assuming that $b/2$ is an integer. This yields a sequence of independent estimators $\{\hat{\beta}_{j,b}^{DC}\}_{j=1}^{B}$. Our estimator $\hat{\beta}_{b}^{DC}$ for $\beta$ is the median of this sequence.

    The algorithm for the \textbf{Divide-and-conquer estimator}  is as follows.
    \setcounter{bean}{0}
\begin{center}
\begin{list}
{\textsc{Step} \arabic{bean}.}{\usecounter{bean}}
\item  Set $B$ such that $n/(2B)$ is an integer.
\item  Partition the data into $B$ adjacent blocks.
\item  Split each block further into two equal sized blocks $R_{j,1}$ and $R_{j,2}$.
\item For $j=1,\ldots,B$, estimate
            \begin{equation}\label{eq:subsampleest}
                \hat{\beta}^{DC}_{j,b} = \frac{\sum_{i\in R_{j,1}}{x}_{i}{y}_{i}^2}{\sum_{i\in R_{j,2}}{x}_{i}^2{y}_{i}}.
            \end{equation}
\item Estimate $\beta$ by
        \begin{equation}
            \hat{\beta}^{DC}_{b} = \text{median}(\hat{\beta}^{DC}_{1,b},\ldots,\hat{\beta}^{DC}_{B,b}).
        \end{equation}
\end{list}
\end{center}

    We prove below that for any fixed $\beta_{0}$ (including $\beta_{0}=0)$,  $\hat{\beta}^{DC}_{b}$ is asymptotically normal. The convergence rate, however, depends on whether $\beta_{0}=0$ or $\beta_{0}\neq 0$. In the first case, the convergence rate is $B^{1/2}$, so it is determined only by the number of blocks $B$. In the latter case, we end up with the standard (pooled) convergence rate of $\sqrt{B\cdot b} = \sqrt{n}$. Hence, the fact that we use the median to estimate $\beta$ is free of any asymptotic costs, as long as the convergence rate of the estimator is concerned.

 \subsection{Selecting the Number of Blocks}
    The choice of the number of subsample estimators $B$ is important. The theoretical results indicate the $B/b$ should go to zero for consistency of the estimator. If $B$ is too large relative to $b$, finite sample bias in the subsample estimators does not wash out over the different blocks. Our simulation results show that we should choose the value of $B$ relatively small. There, we have a total of 60,000 observations and set $B=20$, so that $B/b = 0.0066$ and $B=40$ so that $B/b = 0.0267$. For the latter, we find an increase in bias in the coefficients and confidence intervals that undercover for regressions that include fixed effects. When we increase the cross-section dimension to $n=6,000$ in \cref{table:mcn6000,table:mccov6000}, so that for $B=40$ we get $B/b = 0.0133$ and coverage approaches the nominal rate. Optimal selection of the value of $B$ is an important topic for further research. We suggest that empirical researchers report their estimates for a range of different values of $B$ as a robustness check of their results.

    \subsection{Inference}\label{subsec:inference}
    As the asymptotic variance of the estimator depends on the unknown parameter $\beta_{0}$, we construct confidence intervals from a nonparametric bootstrap procedure. First, we draw $B$ samples $\tilde{e}_{j,b}$ with replacement from $\{\pm e_{1,b},\ldots,\pm e_{B,b}\}$ where $e_{j,b} = \hat{\beta}_{j,b}^{DC}-\hat{\beta}_{b}^{DC}$. Note that in view of the asymptotic distribution, we enforce symmetry by including each $e_{j,b}$ both with a plus and a minus sign. We then calculate $\tilde{\beta}_{i}^{DC}=\text{median}(\tilde{\beta}_{1,b}^{DC},\ldots, \tilde{\beta}_{B,b}^{DC})$, where $\tilde{\beta}_{j,b}^{DC} = \hat{\beta}_{b}^{DC}+\tilde{e}_{j,b}$ and $i=1,\ldots,B_{n}$. The $\alpha/2$ and $1-\alpha/2$ quantiles $(q_{\alpha/2}^{DC},q_{1-\alpha/2}^{DC})$ of the empirical distribution of $\tilde{\beta}_{i}^{DC}-\hat{\beta}_{b}^{DC}$ are then used to construct a confidence interval as $(\hat{\beta}_{b}^{DC}+q_{\alpha/2}^{DC},\hat{\beta}_{b}^{DC}+q_{1-\alpha/2}^{DC})$.

    \subsection{Extensions}
    Two extensions to the above methodology are needed to implement the divide-and-conquer estimator in our empirical setting. The first is the addition of perfectly measured control variables to the model. The second is to consider a panel data setting.

    First, the model \eqref{eq:eiv} can be expanded by including a set of perfectly measured controls, so that the model is of the form
    \begin{equation}
        y_{i} = \xi_{i}\beta_{0} + \bs z_{i}'\bs\gamma + \varepsilon_{i}.
    \end{equation}
    In this case, define
    \begin{equation}
    \dot{x}_{i} = x_{i}-\bs z_{i}'\left(\sum_{i\in R_{j,1}}\bs z_{i}\bs z_{i}'\right)^{-1}\sum_{i\in R_{j,1}}\bs z_{i}x_{i},
    \end{equation}
    and similarly for $\dot{y}_{i}$. Denote by $\ddot{x}_{i}$ and $\ddot{y}_{i}$ the corresponding estimators on the set $R_{j,2}$. We then adapt \eqref{eq:subsampleest} in the divide-and-conquer algorithm to
    \begin{equation}
        \hat{\beta}^{DC}_{j,b} = \frac{\sum_{i\in R_{j,1}}\dot{x}_{i}\dot{y}_{i}^2}{\sum_{i\in R_{j,2}}\ddot{x}_{i}^2\ddot{y}_{i}}.
    \end{equation}
    This retains the independence between the subsample estimators. Note that for inference on $\bs\gamma$ we can adapt the procedure from \cref{subsec:inference} and construct confidence intervals from the quantiles of $\tilde{\bs \gamma}^{DC}_{j,b} = (\sum_{i=1}^{n}\bs z_{i}\bs z_{i}')^{-1}\sum_{i=1}^{n}\bs z_{i}(y_{i}-x_{i}\tilde{\beta}^{DC}_{j,b})$.

    As a second extension, we consider a linear panel data model that will be used in our empirical application,
    \begin{equation}
        y_{it} = x_{it}\beta_{0} +\alpha_{i} + \lambda_{t} +\varepsilon_{it}.
    \end{equation}
    In this case, we construct subsample estimators on cross-sections of the data. To account for individual fixed effects, we perform a within transformation on the data. Moreover, to account for time fixed effects in the model, we demean the data within blocks. Perfectly measured controls are handled in the same way as above.

    The algorithm for the \textbf{Panel Divide-and-conquer estimator} is as follows.
    \setcounter{bean}{0}
\begin{center}
\begin{list}
{\textsc{Step} \arabic{bean}.}{\usecounter{bean}}
\item  Set $B$ such that $N/(2B)$ is an integer.
\item If fixed effects $\alpha_{i}$ are included in the model, then transform $y_{it}$ and $x_{it}$ as
            \begin{equation}
            \widetilde{y}_{it} = y_{it}-T^{-1}\sum_{t=1}^{T}y_{it},\quad \widetilde{x}_{it} = x_{it}-T^{-1}\sum_{t=1}^{T}x_{it}.
            \end{equation}
 \item Partition the cross-section at time $t$ into $B$ blocks;
            \item Split each block further into two equal sized blocks $R_{j,1}$ and $R_{j,2}$\;
            \item For $j=1,\ldots,B$,
               if time effects $\lambda_{t}$ are included in the model, then \\
                    \begin{tabular}{ll}
                        For $i\in R_{j,1}$,&
                        $\widehat{y}_{it} = \widetilde{y}_{it}-(2B/N)\sum_{i\in R_{j,1}}\widetilde{y}_{it}$, \\
                        & $\widehat{x}_{it} = \widetilde{x}_{it}-(2B/N)\sum_{i\in R_{j,1}}\widetilde{x}_{it}$.\\
                        For $i\in R_{j,2}$,&
                        $\widehat{y}_{it} = \widetilde{y}_{it}-(2B/N)\sum_{i\in R_{j,2}}\widetilde{y}_{it}$, \\
                        &$\widehat{x}_{it} = \widetilde{x}_{it}-(2B/N)\sum_{i\in R_{j,2}}\widetilde{x}_{it}$.
                    \end{tabular}\\
            Otherwise set $\widehat{y}_{it}=\widetilde{y}_{it}$ and $\widehat{x}_{it}=\widetilde{x}_{it}$.
 \item  Estimate
\begin{equation}\label{eq:subsampleestpanel}
                    \hat{\beta}^{DC}_{jt,b} = \frac{\sum_{i\in R_{j,1}}\widehat{x}_{it}\widehat{y}_{it}^2}{\sum_{i\in R_{j,2}}\widehat{x}_{it}^2\widehat{y}_{it}}.
                \end{equation}
\item Estimate $\beta$ by
        \begin{equation}
            \hat{\beta}^{DC}_{b} = \text{median}(\hat{\beta}^{DC}_{11,b},\ldots,\hat{\beta}^{DC}_{BT,b}).
        \end{equation}
\end{list}
\end{center}

    \section{Theoretical Results}\label{sec:theory}
    We start by making the following assumption on the moments of $(\xi_{i},u_{i},\varepsilon_{i})$ that facilitate a Berry-Esseen inequality in the proof of the main theorem. We limit our attention to the basic estimator we suggest in Section \ref{section::divide}. In the following, denote by $\Phi(x)$ and $\phi (x)$ the CDF/PDF of a standard normal r.v.\ and by $F_{c}(x)$/$f_{c}(x)$ the CDF/PDF of a Cauchy r.v.

    \begin{assumption}\label{ass:moments}
        $\mathbb{E}[|\xi_{i}|^6]<\infty$, $\mathbb{E}[|u_{i}|^6]<\infty$, $\mathbb{E}[|\varepsilon_{i}|^6]<\infty$.
    \end{assumption}
    We impose the following assumption on the number of blocks $B$, which is chosen as a function of the sample size within each block $b$.
    \begin{assumption}\label{ass:rates}
        $B(b)/b\rightarrow 0$.
    \end{assumption}

    Note that when $\beta_{0}=0$, the estimator can be decomposed as
    \begin{equation}
        \hat{\beta}_{j,b} = \frac{\sum_{i\in R_{j,1}}x_{i}\varepsilon_{i}^2}{\sum_{i\in R_{j,2}}x_{i}^2\varepsilon_{i}} = \frac{v_{j,b}}{w_{j,b}},
    \end{equation}
    where $|R_{j,1}|=|R_{j,2}| =b/2$ and $j=1,\ldots,B$. Denote $\sigma_{v}^2 = \mathbb{E}[v_{j,b}^2]$ and $\sigma_{w}^2=\mathbb{E}[w_{j,b}^2]$. When $\beta_{0}\neq 0$, define $w_{i} = x_{i}\varepsilon_{i}^2 + 2\beta_{0} (\xi_{i}^2 + \xi_{i}u_{i})\varepsilon_{i}$ and $\sigma_{\beta}^2 = (\beta_{0}\sigma_{\xi}^{(3)})^{-2}\mathbb{E}\left[w_{i}^2\right]$.

    This brings us to our main result.

    \begin{theorem}\label{lem:normbeta0} Suppose \Crefrange{ass:errors}{ass:rates} hold. Let $$\hat{\beta}^{DC}_{b} = \emph{median}(\{\hat{\beta}_{j,b}\}_{j=1}^{B}).$$
        (a) When $\beta_{0}=0$,
        \begin{equation}
            \sqrt{B}\left(\frac{\sigma_{w}}{\sigma_{v}}\hat{\beta}_{b}^{DC}\right) \longrightarrow_{d} N\left(0,\frac{1}{4f_{c}(0)^2}\right).
        \end{equation}
        (b) When $\beta_{0}\neq 0$,
        \begin{equation}
            \sqrt{n}\left(\frac{\hat{\beta}_{b}^{DC}-\beta_{0}}{\sigma_{\beta}}\right) \longrightarrow_{d} N\left(0,\frac{1}{4\phi(0)^2}\right).
        \end{equation}
    \end{theorem}
    We see that regardless of the value of $\beta_{0}$, the estimator is asymptotically normal. The asymptotic distribution, on the other hand, is generally discontinuous at $\beta_{0}$. In particular, for $\beta_{0}=0$, the convergence rate is $\sqrt{B}$. Since \cref{ass:rates} requires that $B/b = B^2/n\rightarrow 0$, the maximal convergence rate is below $n^{1/4}$. When $\beta_{0}\neq 0$, we recover the standard $\sqrt{n}$ convergence rate.

    We also see that the asymptotic variance depends on the, evidently unknown, value of $\beta_{0}$. Inference therefore proceeds via the non-parametric bootstrap discussed in the previous section that accommodates cases $(a)$ and $(b)$ simultaneously.

    \section{Numerical Results}\label{sec:numerical}
    \subsection{Monte Carlo Study}
    In this section we consider the simulation set-up from \citet{erickson2014minimum}. Their setup is calibrated in order to reflect several characteristics of the data used in their analysis. In particular, we consider a simple linear panel data model of the form,
    \begin{equation}
        \begin{split}
            y_{it} &=  z_{it}\gamma + \xi_{it}\beta + u_{it},\\
            x_{it} &= \xi_{it} + e_{it}.
        \end{split}
    \end{equation}
    Here $i=1,\ldots,n$ with $n=3,000$ and $t=1,\ldots,T$ with $T=20$. The errors $u_{it}$ and $e_{it}$ are i.i.d.\ Gamma random variables standardized using the population mean and standard deviation. The scale parameter is set to 1 and the shape parameter is set to $0.32$ for $u_{it}$ and $0.09$ for $e_{it}$.

    The perfectly measured controls and the mismeasured regressor are generated using an autoregressive processes of order 1 - AR(1), as follows
    \begin{equation}
        \begin{split}
            \xi_{it} &=\delta_{\xi}+ \phi_{\xi}\xi_{i,t-1} + v_{it}^{\xi},\\
            z_{it}& = \delta_{z} + \phi_{z}z_{i,t-1} + v_{it}^{z}.
        \end{split}
    \end{equation}
    Here we set $\phi_{\xi} = 0.78$ and $\phi_{z}=0.48$. The intercepts are set to match the first moment of Tobin's $q$ and cash flow in the data as $\delta_{\xi} = 0.570$ and $\delta_{z} = 0.094$. The errors $(v_{it}^{\xi},v_{it}^{z})$ are again independent Gamma random variables. The scale parameter is set to 1, the shape parameter for $v_{it}^{\xi}$ equal to 0.007 and for $v_{it}^{z}$ equal to 2.08. We follow the standard practise in the literature and initialize the processes for $\xi_{it}$ and $z_{it}$ in the recent past by setting $\xi_{i,-10}=z_{i,-10}=0$ for all $i$. We then generate $T+10$ time series observations for both processes and drop the first 10 periods.

    \begin{table}[b]
    \caption{}\vspace{-0.25cm}
    \centering
        \caption*{\textit{Simulation: bias and standard deviation.}\vspace{-0.5cm}}
        \begin{center}
        \label{table:mc}
        \setlength{\tabcolsep}{2.6pt}
        \begin{tabular}{@{\extracolsep{4pt}}ll@{\hspace{\tabcolsep}}r@{\hspace{\tabcolsep}}rr@{\hspace{1\tabcolsep}}rr@{\hspace{1\tabcolsep}}rr@{\hspace{1\tabcolsep}}r}
            \hline
            &&\multicolumn{8}{c}{$\beta_{0}=0$, $\alpha_{0}=0.05$}\\
            \cline{3-10}
            &&\multicolumn{2}{c}{(1)}&\multicolumn{2}{c}{(2)}&\multicolumn{2}{c}{(3)}&\multicolumn{2}{c}{(4)}\\
            \cline{3-4}\cline{5-6}\cline{7-8}\cline{9-10}
           OLS&$\widehat{\beta}$&0.000 & (0.000) & 0.000 & (0.000) & 0.000 & (0.000) & 0.000 & (0.000) \\
&$\widehat{\alpha}$&0.050 & (0.002) & 0.050 & (0.002) & 0.050 & (0.002) & 0.050 & (0.002) \\[.5ex]
3M&$\widehat{\beta}$&0.036 & (7.513) & 0.039 & (2.774) & 0.007 & (1.418) & 0.031 & (4.966) \\
&$\widehat{\alpha}$&-0.017 & (14.081) & -0.012 & (4.483) & 0.038 & (2.717) & 0.004 & (7.43) \\[.5ex]
DC(20)&$\widehat{\beta}$&0.001 & (0.008) & 0.001 & (0.008) & 0.001 & (0.007) & 0.001 & (0.008) \\
&$\widehat{\alpha}$&0.048 & (0.015) & 0.049 & (0.013) & 0.048 & (0.014) & 0.049 & (0.013) \\[.5ex]
DC(40)&$\widehat{\beta}$&0.001 & (0.005) & 0.001 & (0.005) & 0.001 & (0.005) & 0.001 & (0.005) \\
&$\widehat{\alpha}$&0.047 & (0.010) & 0.048 & (0.009) & 0.047 & (0.010) & 0.048 & (0.008) \\
             \hline
            &&\multicolumn{8}{c}{$\beta_{0}=0.025$, $\alpha_{0}=0.05$}\\
            \cline{3-10}
            &&\multicolumn{2}{c}{(1)}&\multicolumn{2}{c}{(2)}&\multicolumn{2}{c}{(3)}&\multicolumn{2}{c}{(4)}\\
            \cline{3-4}\cline{5-6}\cline{7-8}\cline{9-10}
            OLS&$\widehat{\beta}$&0.011 & (0.000) & 0.009 & (0.000) & 0.011 & (0.000) & 0.009 & (0.000) \\
&$\widehat{\alpha}$&0.077 & (0.001) & 0.075 & (0.001) & 0.077 & (0.001) & 0.075 & (0.001) \\[.5ex]
3M&$\widehat{\beta}$&0.025 & (0.000) & 0.025 & (0.001) & 0.025 & (0.000) & 0.025 & (0.001) \\
&$\widehat{\alpha}$&0.050 & (0.002) & 0.050 & (0.002) & 0.050 & (0.002) & 0.050 & (0.002) \\[.5ex]
DC(20)&$\widehat{\beta}$&0.026 & (0.007) & 0.024 & (0.01) & 0.026 & (0.007) & 0.025 & (0.01) \\
&$\widehat{\alpha}$&0.048 & (0.014) & 0.051 & (0.015) & 0.048 & (0.014) & 0.050 & (0.015) \\[.5ex]
DC(40)&$\widehat{\beta}$&0.026 & (0.007) & 0.018 & (0.007) & 0.026 & (0.007) & 0.019 & (0.008) \\
&$\widehat{\alpha}$&0.049 & (0.013) & 0.061 & (0.011) & 0.049 & (0.013) & 0.059 & (0.012) \\
            \hline
            FE && \multicolumn{2}{c}{No} &  \multicolumn{2}{c}{Yes} &  \multicolumn{2}{c}{No} &  \multicolumn{2}{c}{Yes} \\
            TE && \multicolumn{2}{c}{No} &  \multicolumn{2}{c}{No} &  \multicolumn{2}{c}{Yes} &  \multicolumn{2}{c}{Yes} \\
               \hline
        \end{tabular}\vspace{-0.25cm}
        \end{center}
        \begin{minipage}{0.98\textwidth}
            \textit{Note:} mean and standard deviation of the estimates for $\beta$ and $\alpha$ over $20,000$ draws of the error-in-variables model $y_{it} = \xi_{it}\beta + z_{it}\alpha+u_{i}$, $x_{it} = \xi_{it}+v_{it}$ with $n=3,000$ and $T=20$. Estimation by least squares (OLS), the third-order moment estimator by \citet{geary1941inherent} (3M), and the divide-and-conquer estimator with $B$ blocks ($\text{DC}_{B}$). The standard deviation of the estimates is listed in brackets. Model (1) includes an intercept, model (2) fixed effects, model (3) time effects, and model (4) two-way fixed effects. For DC, in the presence of fixed effects, we do a within transformation on the complete data set. For time effects, we demean within blocks.
        \end{minipage}
    \end{table}

    After having generated $z_{it}$ and $\xi_{it}$, we transform these to ensure that the covariance matrix calculated using all available data is equal to the covariance matrix of $x_{it}$ and $z_{it}$ in the data, with the variance of Tobin's $q$ multiplied by $\tau^2 = 0.45$ and the covariances multiplied by $\sqrt{\tau^2}$. Numerically, the covariance matrix is
    \begin{equation}
        C = \left(\begin{array}{cc}
            16.130 & \\
            0.489 & 0.258
        \end{array}\right).
    \end{equation}
    Finally, we generate the observed variables as follows
    \begin{equation}
        \begin{split}
            x_{it} &= \xi_{it} + \sqrt{\tau^{-2}(1-\tau^2)C_{1,1}}e_{it}\\
            y_{it}  &= \mu_{y} + \xi_{it}\beta + z_{it}\gamma + \sqrt{\sigma_{y}^2-\text{var}(\xi_{it}\beta+z_{it}\gamma)}u_{it}\label{dgp}.
        \end{split}
    \end{equation}

    Here $\mu_{y}$ and $\sigma_{y}^2$ are set to match the population mean and variance of the investment variable in the data. We consider two choices for $\beta$. In the first $\beta=0.025$ as in \citet{erickson2014minimum}, while in the second $\beta=0$. The coefficient for control variable $z_{it}$ is fixed to $\gamma=0.05$.

    We estimate the parameters in \eqref{dgp} using OLS (OLS), the third-order moment estimator (3M) and the divide-and-conquer estimator (DC). For the divide-and-conquer estimator, we vary the number of blocks per time period from 1 to 2, so we take the median over $B\cdot T = 20$ and $B\cdot T = 40$ subsample estimators. In each time period, the partition of the firms is random.

    Bootstrap confidence intervals are constructed as outlined in \cref{subsec:inference} with $B_{n}=399$. For the third-order moment estimator, confidence intervals are calculated as in \citet{erickson2002two}.

    \begin{table}[b]
        \caption{}\vspace{-0.25cm}
        \centering
        \caption*{\textit{Simulation: coverage.}\vspace{-0.5cm}}
        \begin{center}
        \label{table:mccov}
        \begin{tabular}{@{\extracolsep{4pt}}llrrrrrrrr}
            \hline
            &&\multicolumn{4}{c}{$\beta_{0}=0$, $\alpha_{0}=0.05$}&\multicolumn{4}{c}{$\beta_{0}=0.025$, $\alpha_{0}=0.05$}\\
            \cline{3-6}\cline{7-10}
           OLS &$\beta$&  95.2 & 94.5 & 95.1 & 94.5 & 0.0 & 0.0 & 0.0 & 0.0 \\
 &$\alpha$&  95.1 & 94.4 & 95.1 & 94.3 & 0.0 & 0.0 & 0.0 & 0.0 \\[.5ex]
3M &$\beta$&  96.6 & 98.2 & 96.6 & 98.2 & 94.7 & 95.5 & 94.8 & 95.5 \\
 &$\alpha$& 96.5 & 98.3 & 96.5 & 98.2 & 95.0 & 94.7 & 95.0 & 94.7 \\[.5ex]
DC(20) &$\beta$& 96.6 & 97.0 & 96.7 & 96.9 & 94.1 & 92.3 & 93.9 & 92.7 \\
 &$\alpha$&  96.7 & 96.7 & 96.5 & 96.7 & 93.9 & 92.2 & 93.7 & 92.7 \\[.5ex]
DC(40) &$\beta$&  95.7 & 96.3 & 95.5 & 96.3 & 94.3 & 75.5 & 94.2 & 79.8 \\
 &$\alpha$& 95.4 & 96.1 & 95.4 & 95.9 & 94.0 & 75.3 & 94.1 & 79.8 \\
            \hline
            FE && \multicolumn{1}{c}{No} & \multicolumn{1}{c}{Yes} & \multicolumn{1}{c}{No} & \multicolumn{1}{c}{Yes}& \multicolumn{1}{c}{No} & \multicolumn{1}{c}{Yes} & \multicolumn{1}{c}{No} & \multicolumn{1}{c}{Yes}\\
            TE && \multicolumn{1}{c}{No} & \multicolumn{1}{c}{No} & \multicolumn{1}{c}{Yes} & \multicolumn{1}{c}{Yes}& \multicolumn{1}{c}{No} & \multicolumn{1}{c}{No} & \multicolumn{1}{c}{Yes} & \multicolumn{1}{c}{Yes}\\
            \hline
        \end{tabular}\vspace{-0.25cm}
        \end{center}
        \begin{minipage}{0.93\textwidth}
       \textit{Note:} reported is the coverage rate ($\times 100$) for the model and methods as describe below \cref{table:mc}. Nominal coverage rate is 0.95. Confidence intervals for the 3M estimator are calculated as in \citet{erickson2002two}. Confidence intervals for the DC estimator are obtained by a bootstrap procedure. To obtain a bootstrap draw, we draw samples $e_{i,b}^{*}$ with $i=1,\ldots,B$ with replacement from $\{\pm e_{1,b},\ldots,\pm e_{B,b}\}$ where $e_{i,b} = \hat{\beta}_{i,b}-\hat{\beta}_{b}^{DC}$ and calculate the median of $\{e_{1,b}^{*} + \hat{\beta}^{DC}_{b},\ldots, e_{B,b}^{*}+\hat{\beta}_{b}^{DC}\}$. We repeat this procedure 399 times and form a confidence interval by taking the 5 and 95 percentiles.
       \end{minipage}
    \end{table}

    \begin{center}
        [\cref{table:mc} near here.]
    \end{center}
    \cref{table:mc} shows the mean and standard deviations of the estimates for the $q$-coefficient $\beta$ and the cash flow coefficient $\alpha$. When $\beta_{0}=0.025$, we see that the OLS estimates for $\beta$ and $\alpha$ are badly biased due to the measurement error in the process. The third-order moment estimators as well as the divide-and-conquer methods show no to very little bias. The standard errors of the divide-and-conquer estimators are larger relative to the third-order moment estimator.

    When $\beta_{0}=0$, we see that the third-order moment estimator is badly defined, with the standard error going up substantially.  The divide-and-conquer estimators on the other hand perform well, with standard deviations nearly identical to those when $\beta_{0}=0.025$.

     \begin{center}
        [\cref{table:mccov} near here.]
    \end{center}

    \cref{table:mccov} shows the coverage rates for the various methods. As expected, the bias in the OLS estimator when $\beta_{0}\neq 0$ results in confidence intervals that do not include the true parameter. The confidence intervals corresponding to the 3M estimator are somewhat conservative, especially when we include fixed effects. For the DC estimator, we find overall close to nominal coverage except when $\beta_{0}=0.025$ and we include fixed effects. \cref{table:mccov6000} shows that increasing the sample size to $n=6,000$ largely resolves this issue.

    \subsection{Empirical Results}
    We consider an application from the corporate finance literature that seeks to determine the relation between actual investment and unobserved investment opportunities in capital as in \citet{fazzari1988financing}. The investment opportunities are proxied by Tobin's $q$, which contains considerable measurement error. The third-order moment estimator \eqref{eq:gmm} and generalizations thereof have been applied to account for this measurement error in \citet{erickson2000measurement}, \citet{erickson2012treating} and \citet{erickson2014minimum}. While identification problems due to lack of skewness were investigated in \citet{erickson2012treating}, problems that arise because $\beta_{0}$ is (close to) zero, have not been addressed.

     \begin{table}[b]
     \caption{}\vspace{-.25cm}
     \centering
        \caption*{\textit{Application: full sample estimates.}\vspace{-0.5cm}}
        \begin{center}
        \label{tab:fullsample}
        \setlength{\tabcolsep}{2.6pt}
        \begin{tabular}{@{\extracolsep{4pt}}lcccc}
            \hline
            & \multicolumn{1}{c}{(1)} &\multicolumn{1}{c}{(2)} &\multicolumn{1}{c}{(3)} &\multicolumn{1}{c}{(4)} \\
            \cline{2-2}\cline{3-3}\cline{4-4}\cline{5-5}
            OLS & 0.008 &   0.009 & 0.009   & 0.009  \\
 & (0.008, 0.008) & (0.009, 0.009) & (0.009, 0.009) & (0.008, 0.009) \\[.5ex]
3M & 0.024 & 0.041 & 0.024 & 0.037   \\
 & (0.022, 0.025) & (0.034, 0.048) & (0.023, 0.026) & (0.031, 0.042) \\[.5ex]
DC(2) & 0.023 & 0.031 & 0.035 & 0.020   \\
 & (0.013, 0.034) & (0.015, 0.048) & (0.023, 0.048) & (-0.005, 0.045) \\[.5ex]
DC(1) & 0.035 & 0.035 & 0.040 & 0.030   \\
 & (0.023, 0.047) & (0.015, 0.056) & (0.030, 0.050) & (-0.002, 0.062) \\
\hline
OLS & 0.059 & 0.095 & 0.052 & 0.091   \\
 & (0.056, 0.061) & (0.093, 0.098) & (0.05, 0.054) & (0.089, 0.094) \\[.5ex]
3M & 0.011 & -0.025 & 0.003 & -0.013   \\
 & (0.004, 0.018) & (-0.051, 0.002) & (-0.004, 0.010) & (-0.035, 0.009) \\[.5ex]
DC(2) & 0.012 & 0.014 & -0.033 & 0.049   \\
 & (-0.019, 0.043) & (-0.048, 0.084) & (-0.075, 0.008) & (-0.033, 0.123) \\[.5ex]
DC(1) & -0.024 & -0.003 & -0.049 & 0.010   \\
 & (-0.059, 0.011) & (-0.08, 0.074) & (-0.081, -0.015) & (-0.112, 0.136) \\
            \hline
            FE & No & Yes & No & Yes\\
            TE & No & No & Yes & Yes\\
            \hline
        \end{tabular}\vspace{-0.25cm}
        \end{center}
        \begin{minipage}{0.95\textwidth}
        \textit{Note:} Top panel: estimates for the effect of Tobin's $q$ on investment. Bottom panel: estimates for the effect of cash flow on investment. 95\% confidence intervals in brackets. DC($B$) indicates the divide-and-conquer estimator with $B$ blocks per year.
        \end{minipage}
    \end{table}

    We use the data of \citet{erickson2014minimum} to estimate the  linear panel data model
    \begin{equation}\label{eq:modelapplication}
        y_{it} = \bs z_{it}'\bs\gamma + x_{it}\beta + \varepsilon_{it},
    \end{equation}
    where $i=1,\ldots,n$ indexes the firm, and $t=1,\ldots,T$ the year. The data ranges from 1970 to 2011. After removing firms for which only one year is available, the sample consists of 121,733 firm-year observations. The outcome variable $y_{it}$ is investment as constructed by \citet{erickson2014minimum}. The perfectly measured control variables $\bs z_{it}$ include an intercept and a measure of cash flow, $x_{it}$ denotes Tobin's $q$.

    We estimate the effect of Tobin's $q$ on investment in \eqref{eq:modelapplication} by (1) OLS, (2) the third-order moment estimator \eqref{eq:gmm} and (3) the divide-and-conquer algorithm. We calculate confidence intervals for \eqref{eq:gmm} as described in \citet{erickson2002two}. For the divide-and-conquer method, we apply the bootstrap procedure from \cref{subsec:inference}.

    For the divide-and-conquer estimator, for every year we divide the data into 2 blocks, which for the full sample corresponds to 80 blocks. We also report results for 1 block per year. Since for each block we split the data into two halves to calculate the numerator and denominator of the subsample estimators, in each year we discard a randomly selected subset of firms so that the number of firms is a multiple of 2 (for DC with 1 block per year) or a multiple of 4 (for DC with 2 blocks per year).

     \begin{center}
        [\cref{tab:fullsample} near here.]
    \end{center}

    We first report the full sample pooled estimators and standard errors in \cref{tab:fullsample}. For the OLS estimators, we see the significant effect of cash flow on investment. This questions $q$-theory, which states that in a regression of investment on $q$, control variables must be insignificant. Indeed, we see that this theory can be aligned with empirical evidences by the third-order moment estimators. These conclusions are upheld by the divide-and-conquer methods: Tobin's $q$ has a significant positive effect on investment, while cash flow does not. The exception is formed by the results from divide-and-conquer with both firm and year effects. In this case, the confidence intervals for the Tobin's $q$ include zero.

    Secondly, \citet{erickson2014minimum} document that the coefficient estimates may be varying over time. We therefore apply the same estimation method on subsamples of the data. The first subsample is 1970-1980, which we then expand by one year at a time until we reach the full-sample estimates in 2011. For this exercise, we limit our attention to the model with firm fixed effects. We report results for the divide-and-conquer estimator with $2$ blocks per year. Results for $1$ block per year are reported in \cref{fig:tobinsqB1,fig:cashflowB1}.

\begin{figure}[b]
        \caption{Application: estimates for the effect of Tobin's $q$ on investment.}\vspace{-0.5cm}
        \begin{center}
        \label{fig:tobinsq}
       \includegraphics[width=\textwidth]{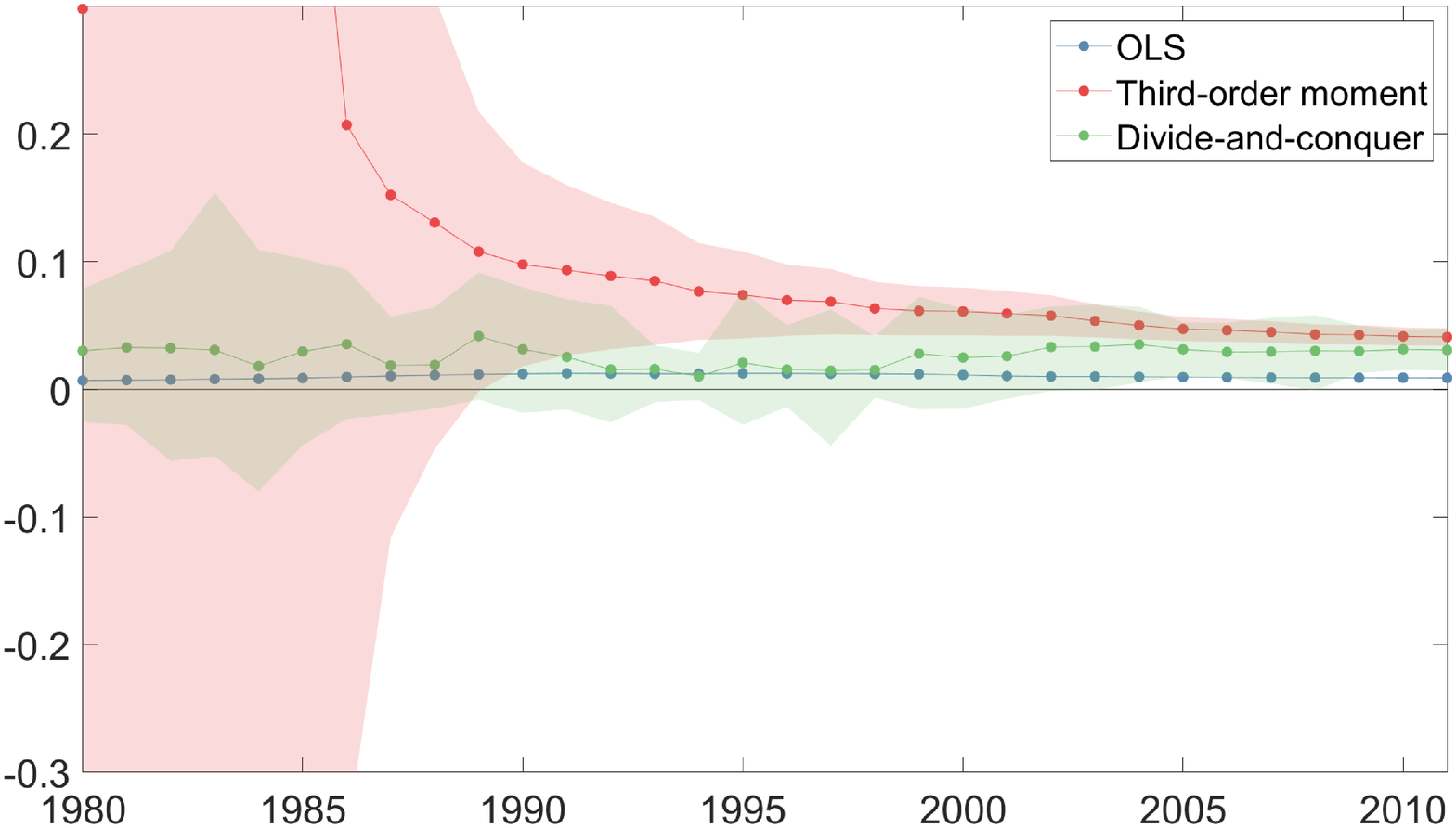}
       \end{center}
    \end{figure}

    \begin{center}
        [\cref{fig:tobinsq} near here.]
    \end{center}

    \cref{fig:tobinsq} shows the estimates for Tobin's $q$ alongside the 95\% confidence intervals. Interestingly, in the first half of the time period, we find that both the third-order moment estimator and the DC estimator include zero in the confidence interval. This is paired with the observation that the third-order moment estimator takes a value for Tobin's $q$ which is an order of magnitude larger than the full-sample estimator.

\begin{figure}[b]
    \centering
        \caption{Application: estimates for the effect of cash flow on investment.}\vspace{-0.5cm}
        \begin{center}
        \label{fig:cashflow}
        \includegraphics[width=\textwidth]{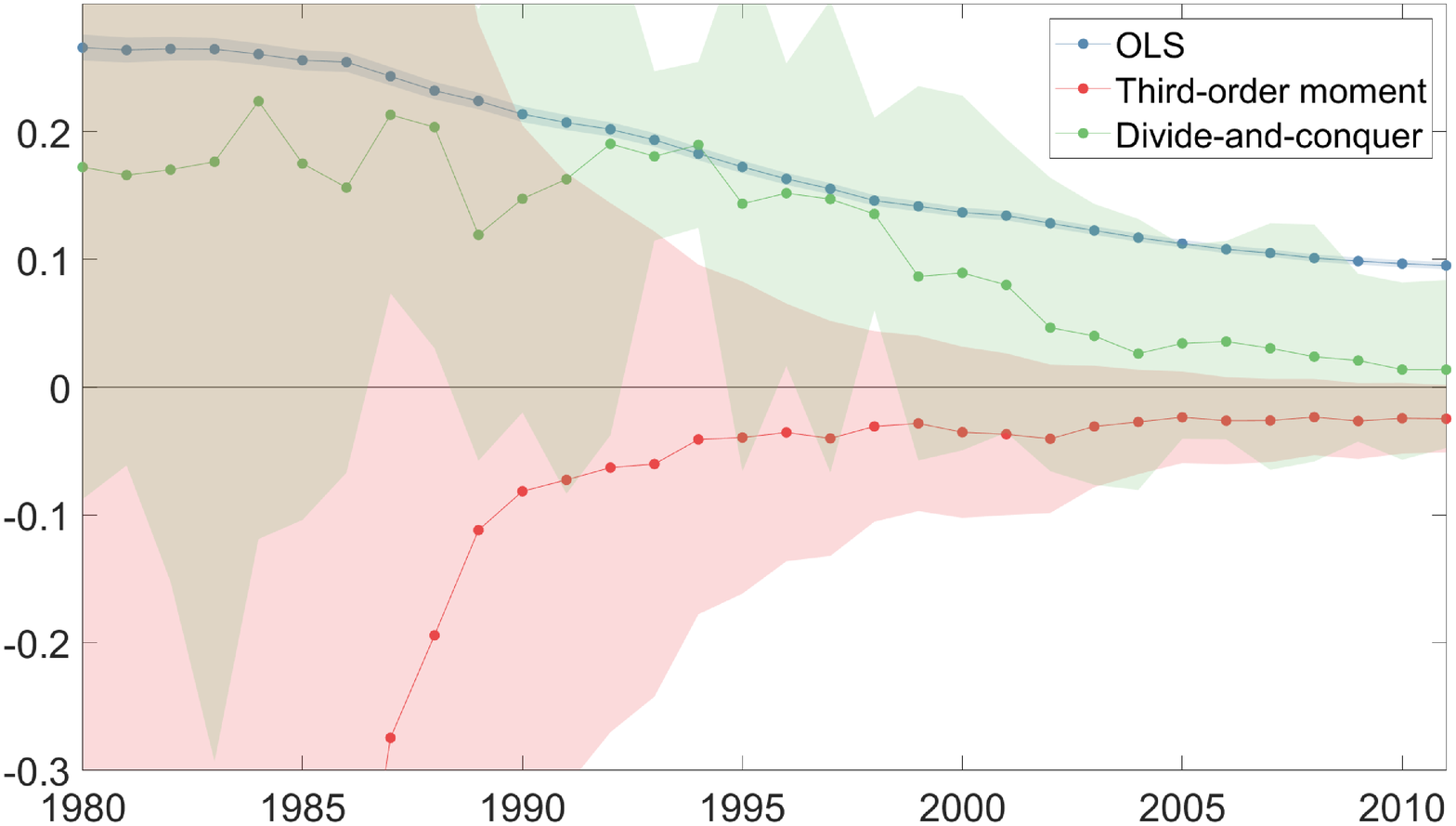}
        \end{center}
      \end{figure}

    \begin{center}
        [\cref{fig:cashflow} near here.]
    \end{center}

    \cref{fig:cashflow} shows the estimated cash flow coefficient. While the confidence intervals for the third-order moment estimator include zero for all time periods, the divide-and-conquer estimator shows some mild evidence for a cash flow effect, especially in the early years of the sample. This aligns with the findings by \citet{andrei2019did} who find that $q$-theory has become increasingly efficient in explaining investment patterns.

    \section{Concluding Remarks}\label{sec:conclusion}
Estimators based on higher-order moments of the data can eliminate the effect of measurement error. However, if the coefficient $\beta_{0}$ of the latent regressor is zero, the estimators converge weakly to a ratio of correlated mean zero normal random variable. As a result, standard inference procedures are not applicable in this case.

As a solution to this problem we suggest the divide-and-conquer estimator that estimates the parameter of interest using the median of subsample estimators. We show this estimator is consistent and suggest a bootstrap based inference procedure that is applicable irrespective of the coefficient $\beta_{0}$. The estimator and inference procedure are intuitive and easy to implement by practitioners. Monte Carlo results indicate that this bootstrap procedure well approximates the finite sample distribution of the estimator.

Finally, we analyse the data on firm investment, Tobin's $q$ and cash flow from \citet{erickson2014minimum} and do not find significant evidence for the effect of cash flow in firm investment. Interestingly enough, we find substantial changes in the estimated relation between investment, Tobin's $q$ and cash flow over time. In particular, we find rather extreme point estimates and confidence intervals in time periods before the mid-1980s for the third-moment estimator of \citet{erickson2014minimum}. These occur precisely in time periods in which the divide-and-conquer estimator is not significantly different from zero. On the other hand, the divide-and-conquer estimates for the effect of Tobin's $q$ is found to be rather stable over time. These observations confirm the empirical relevance of our proposed statistical procedure.

\paragraph{Acknowledgements}
We thank Gerard van den Berg, Simon Broda, Noud van Giersbergen, Toru Kitagawa, Frank Kleibergen, Ruud Koning, Tom Wansbeek, and seminar participants at the Tinbergen Institute Amsterdam and the University of Groningen, for helpful comments. We thank Toni Whited for sharing the data. Financial support from the Netherlands Organization for Scientific Research (NWO) under research grant number $201\text{E}.011$ (TB) and $451-17-002$ (AJ) is gratefully acknowledged.

\paragraph{Disclosure statement}
The authors report there are no competing interests to declare.

\paragraph{Data availability statement}
The authors confirm that the code and data supporting the findings of this study are available within the supplementary materials.
\setlength{\bibsep}{0pt plus 0.2ex}
\bibliographystyle{apalike}
    \bibliography{literature}

\newpage
 \begin{appendices}
        \crefalias{section}{appsec}
        \crefalias{subsection}{appsubsec}
        \numberwithin{equation}{section}
        \section{Proofs}\label{app:proofs}
        \subsection{The Case when \texorpdfstring{$\beta_{0}=0$}{beta = 0}}

        Denote the subsample estimator
        \begin{equation}
            \hat{\beta}_{j,b}^{DC}= \frac{\sum_{i\in R_{j,1}}x_{i}\varepsilon_{i}^2}{\sum_{i\in R_{j,2}}x_{i}^2\varepsilon_{i}} = \frac{v_{j,b}}{w_{j,b}},
        \end{equation}
        where $|R_{j,1}|=|R_{j,2}| =b/2$ and $j=1,\ldots,B$. Let $\sigma_{v}^2 = \mathbb{E}[v_{j,b}^2]$ and $\sigma_{w}^2=\mathbb{E}[w_{j,b}^2]$ and denote $\eta = \sigma_{w}/\sigma_{v}$.
        Denote $\hat{F}(x)=\frac{1}{B}\sum_{j=1}^{B}1[\eta\cdot \hat{\beta}_{j,b}^{DC}\leq x]$ and $F(x)=\mathbb{P}\left(\eta\cdot \hat{\beta}_{j,b}^{DC} \leq x\right)$. Denote by $F_{c}(x)$ and $f_{c}(x)$ the CDF and PDF of a Cauchy random variable.

        \subsubsection{Subsample Estimator Distribution}

        \begin{lemma}\label{lem:berry} If $\beta_{0}=0$, we have that
            \begin{equation}
                \sup_{x\in\mathbb{R}}\left|F(x)-F_{c}(x)\right| = O(b^{-1/2}).
            \end{equation}
        \end{lemma}
        \noindent \textit{Proof}: We have that
        \begin{equation}
            \begin{split}
                F(x)=\mathbb{P}\left(\eta\cdot \hat{\beta}_{j,b}^{DC} \leq x\right)  &= \mathbb{P}\left(v_{j,b}/\sigma_{v}\leq x\cdot w_{j,b}/\sigma_{w},w_{j,b}>0\right) \\
                &\qquad +\mathbb{P}\left(v_{j,b}/\sigma_{v}\geq  x\cdot w_{j,b}/\sigma_{w},w_{j,b}<0\right).
            \end{split}
        \end{equation}
        For $k=1,2$, define $V_{j,b} = b^{-1/2}v_{j,b}/\sigma_{v}$  and $W_{j,b} = b^{-1/2}w_{j,b}/\sigma_{w}$ Under \cref{ass:moments}, by a standard Berry-Esseen bound, $\sup_{x\in \mathbb{R}}\left|\mathbb{P}(V_{j,b}\leq x) - \Phi(x)\right|=O(b^{-1/2})$ and $\sup_{x\in \mathbb{R}}\left|\mathbb{P}(W_{j,b}\leq x) - \Phi(x)\right|=O(b^{-1/2})$.
        Then also,
        \begin{equation}
            \begin{split}
                F(x) &=\int_{0}^{\infty}f_{W_{j,b}}(y)\int_{-\infty}^{xy}f_{V_{j,b}}(z)dzdy+ \int_{-\infty}^{0}f_{W_{j,b}}(y)\int_{xy}^{\infty}f_{V_{j,b}}(z)dzdy\\
                &=\int_{0}^{\infty}\phi(y)\Phi(xy)dy + \int_{-\infty}^{0}\phi(y)(1-\Phi(xy))dy + O(b^{-1/2}),
            \end{split}
        \end{equation}
        and the result holds uniformly over $x$. Here, for simplicity, we assume that the underlying variables $V_{j,b}$ hand $W_{j,b}$ have a continuous distribution. We make this assumption only for the sake of exposition as the main result evidently follows even without the continuity.

        Note that for any two independent standard normal random variables, likewise
        \begin{equation}
            \begin{split}
                \mathbb{P}\left(Z_{1}/Z_{2} \leq x\right) =\int_{0}^{\infty}\phi(y)\Phi(xy)dy + \int_{-\infty}^{0}\phi(y)(1-\Phi(xy))dy.
            \end{split}
        \end{equation}
        and the result follows.\hfill$\blacksquare$

        \subsubsection{Proof for Theorem 1 - Part (a)}

        Suppose for simplicity that $B$ is odd. Denote by $T_{B}=\sum_{j=1}^{B}I[\eta\cdot \hat{\beta}_{j,b}^{DC}>t/\sqrt{B}]$. Since $\hat{\beta}^{DC}_{b}=\text{median}(\hat{\beta}_{1,b}^{DC},\ldots,\hat{\beta}_{B,b}^{DC})$, we have that
        \begin{equation}\label{eq:TB}
            \mathbb{P}(\eta\cdot \hat{\beta}_{b}^{DC}\leq t/\sqrt{B}) = \mathbb{P}(T_{B}\leq (B-1)/2).
        \end{equation}
        Because the estimators $\hat{\beta}_{j,b}^{DC}$ are calculated on independent samples, we have that $T_{B}\sim \text{Bin}(B,p_{B}(t))$ with $p_{B}(t)=1-F(t/\sqrt{B}) = 1-F_{c}(t/\sqrt{B}) + O(1/\sqrt{b})$ by \cref{lem:berry}. Now, the r.h.s.\ of \eqref{eq:TB} equals
        \begin{equation}
            \mathbb{P}\left(\frac{T_{B}-Bp_{B}(t)}{\sqrt{Bp_{B}(t)(1-p_{B}(t))}}\leq \frac{(B-1)/2-Bp_{B}(t)}{\sqrt{Bp_{B}(t)(1-p_{B}(t))}}\right) = \mathbb{P}\left(\sum_{j=1}^{B(b)}Y_{j,b}\leq x_{B}(t)\right).
        \end{equation}
        We have that
        \begin{equation}
            \begin{split}
                x_{B}(t)&=\frac{(B-1)/2-Bp_{B}(t)}{\sqrt{Bp_{B}(t)(1-p_{B}(t))}} \\
                &= \frac{\sqrt{B}(1-p_{B}(t)-1/2)+O(1/\sqrt{B})}{\sqrt{p_{B}(t)(1-p_{B}(t))}}\\
                &=\frac{\sqrt{B}\left(F_{c}(t/\sqrt{B})-F_{c}(0)\right) + O(\sqrt{B/b}) + O(1/\sqrt{B})}{\sqrt{(1-F_{c}(t/\sqrt{B}))F_{c}(t/\sqrt{B}) + O(1/\sqrt{b})}}\\
                &=t\frac{F_{c}(t/\sqrt{B})-F_{c}(0)}{(t/\sqrt{B})\sqrt{1/4 +O(1/\sqrt{B}) + O(1/\sqrt{b})}}+o(1)\\
                & =2f_{c}(0)t + o(1).
            \end{split}
        \end{equation}
        We apply a standard CLT for triangular arrays to $\sum_{j=1}^{B(b)}Y_{j,b}$ to find that
        \begin{equation}
            2\eta f_{c}(0)\cdot \sqrt{B}(\hat{\beta}^{DC}_{b}-0)\longrightarrow_{d} N(0,1).
        \end{equation}

        \subsection{The Case when \texorpdfstring{$\beta_{0}\neq 0$}{beta =/= 0}}
        We first define the following: $\hat{F}(x)=\frac{1}{B}\sum_{j=1}^{B}1[\sqrt{b}(\hat{\beta}_{j,b}^{DC}-\beta_{0})/\sigma_{\beta}\leq x]$ and $F(x)=\mathbb{P}\left(\sqrt{b}(\hat{\beta}_{j,b}^{DC}-\beta_{0})/\sigma_{\beta}\leq x\right)$. We write out $\hat{\beta}_{j,b}^{DC}$ as
        \begin{equation}
            \begin{split}
                \hat{\beta}_{j,b}^{DC}& = \frac{\sum_{i\in R_{j,1}}\left(\xi_{i}^3\beta_{0}^2 + x_{i}\varepsilon_{i}^2 + 2\beta_{0} (\xi_{i}^2 + \xi_{i}u_{i})\varepsilon_{i}\right)}{\sum_{i\in R_{j,2}}\left(\xi_{i}^3\beta_{0} + x_{i}^2\varepsilon_{i} + \beta_{0} (u_{i}^2+2u_{i}\xi_{i})\xi_{i}\right)}\\
                &=\beta +  \frac{\frac{1}{b}\sum_{i\in R_{j,1}}\left(x_{i}\varepsilon_{i}^2 + 2\beta_{0} (\xi_{i}^2 + \xi_{i}u_{i})\varepsilon_{i}\right)}{\beta \sigma_{\xi}^{(3)}} + O_{p}(b^{-1}).
            \end{split}
        \end{equation}
        Define $w_{i} = x_{i}\varepsilon_{i}^2 + 2\beta_{0} (\xi_{i}^2 + \xi_{i}u_{i})\varepsilon_{i}$ and $\sigma_{\beta}^2 = (\beta_{0}\sigma_{\xi}^{(3)})^{-2}\mathbb{E}\left[w_{i}^2\right]$, then
        \begin{equation}
            \mathbb{P}\left(\sqrt{b}\frac{(\hat{\beta}_{j,b}^{DC}-\beta_{0})}{\sigma_{\beta}}\leq x\right)  = \mathbb{P}\left(\frac{\frac{1}{\sqrt{b}}\sum_{i\in R_{j,1}}w_{i}}{\mathbb{E}[w_{i}^2]^{1/2}} \leq x\right) +O(b^{-1}).
        \end{equation}
        Using again a Berry-Esseen bound facilitated by \cref{ass:moments}, we get
        \begin{equation}
            \sup_{x\in \mathbb{R}}\left|F(x)-\Phi(x)\right| = O(1/\sqrt{b}).
        \end{equation}
        \subsubsection{Proof for Theorem 1 - Part (b)}
\begin{figure}[b]
        \caption{Application: estimates for the effect of Tobin's $q$ on investment (1 block per year). }
        \label{fig:tobinsqB1}
        \begin{center}
        \includegraphics[width=\textwidth]{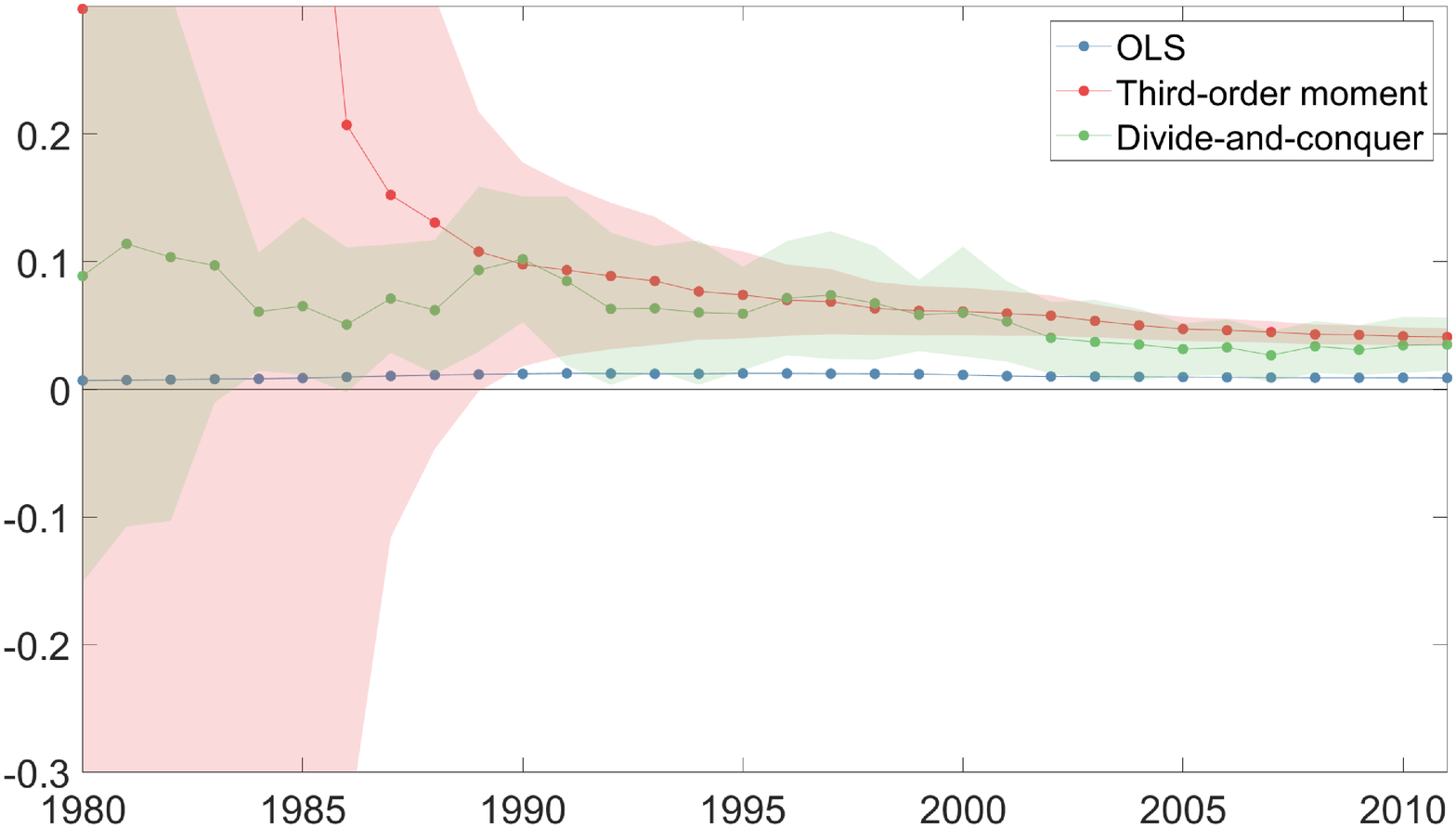}
        \end{center}
        \end{figure}

\begin{figure}[b]
        \caption{Application: estimates for the effect of cash flow on investment (1 block per year). }
        \label{fig:cashflowB1}
        \begin{center}
        \includegraphics[width=\textwidth]{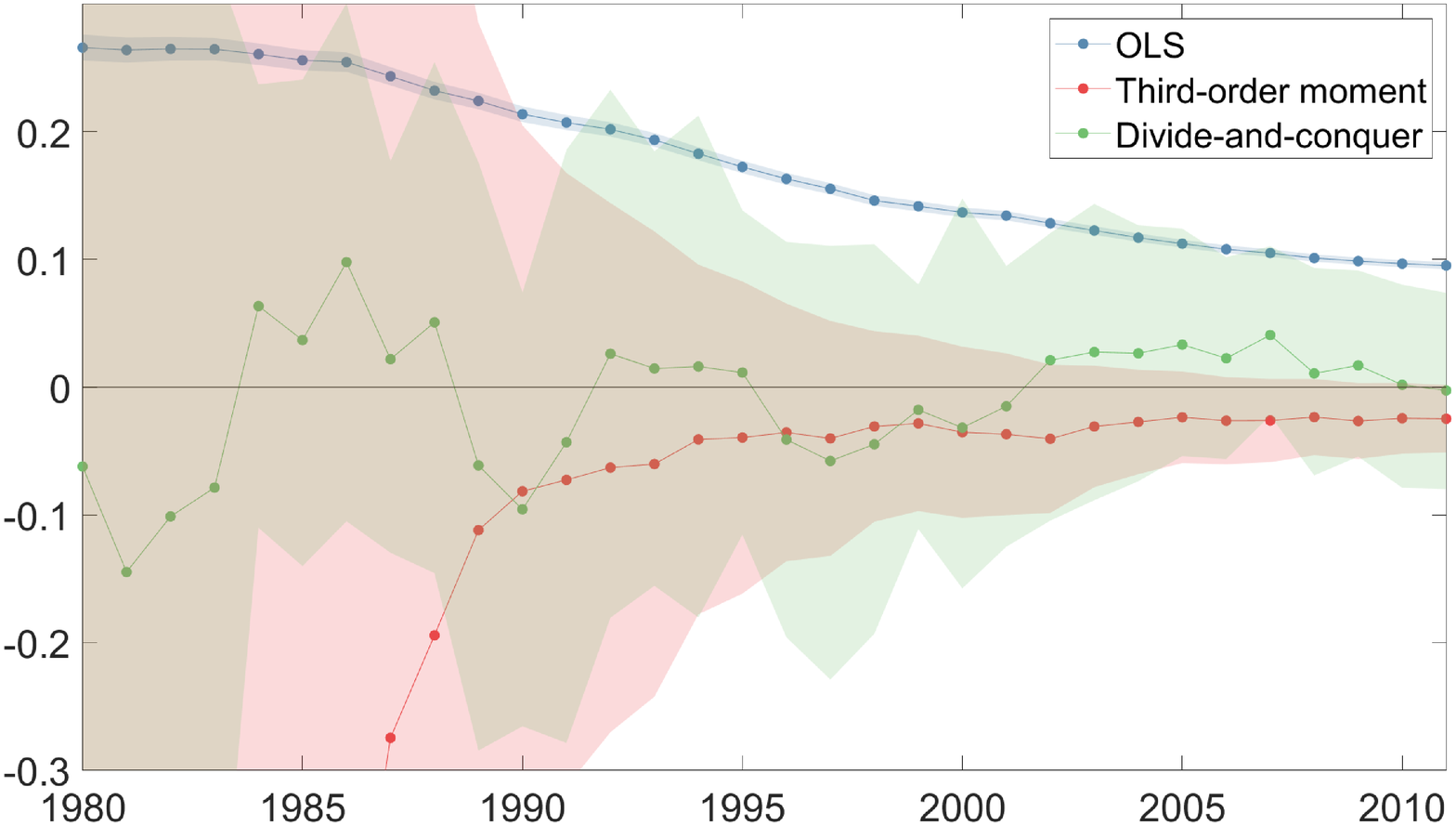}
        \end{center}
    \end{figure}

        The proof is similar to that of \cref{lem:normbeta0}. Suppose again that $B$ is odd. Denote by $T_{B}=\sum_{j=1}^{B}I[\sqrt{b}(\hat{\beta}_{j,b}^{DC}-\beta)/\sigma_{\beta}>t/\sqrt{B}]$, then
        \begin{equation}\label{eq:TB2}
            \mathbb{P}\left(\sqrt{b}(\hat{\beta}^{DC}-\beta)/\sigma_{\beta}\leq t/\sqrt{B}\right) = \mathbb{P}(T_{B}\leq (B-1)/2).
        \end{equation}
        Now $T_{B}\sim \text{Bin}(B,p_{B})$ with $p_{B}=1-F(t/\sqrt{B}) = 1-\Phi(t/\sqrt{B}) + O(1/\sqrt{b})$. Proceeding as in the proof for \cref{lem:normbeta0}, the r.h.s.\ of \eqref{eq:TB2} equals
        \begin{equation}
            \mathbb{P}\left(\frac{T_{B}-Bp_{B}}{\sqrt{Bp_{B}(1-p_{B})}}\leq \frac{(B-1)/2-Bp_{B}}{\sqrt{Bp_{B}(1-p_{B})}}\right) = \mathbb{P}\left(\sum_{i=1}^{B(b)}Y_{i,b}\leq x_{B}\right).
        \end{equation}
        Here
        \begin{equation}
            \begin{split}
                x_{B}&=\frac{(B-1)/2-Bp_{B}}{\sqrt{Bp_{B}(1-p_{B})}}\\
                &= \frac{\sqrt{B}(1-p_{B}-1/2)+O(1/\sqrt{B})}{\sqrt{p_{B}(1-p_{B})}}\\
                &=\frac{\sqrt{B}\left(\Phi(t/\sqrt{B})-\Phi(0)\right) + O(\sqrt{B/b}) + O(1/\sqrt{B})}{\sqrt{(1-\Phi(t/\sqrt{B}))\Phi(t/\sqrt{B}) + O(1/\sqrt{b})}}\\
                &=t\frac{\Phi(t/\sqrt{B})-\Phi(0)}{(t/\sqrt{B})\sqrt{1/4 +O(1/\sqrt{B}) + O(1/\sqrt{b})}}+o(1)\\
                & = 2t\phi(0) + o(1).
            \end{split}
        \end{equation}
        We now find that
        \begin{equation}
            2\sigma_{\beta}\phi(0)\cdot \sqrt{B\cdot b}(\hat{\beta}^{DC}-\beta_{0})\longrightarrow_{d}N(0,1).
        \end{equation}
        \newpage
        \begin{center}
        [\cref{table:mcn6000} near here.]
        \end{center}
        \begin{center}
         [\cref{table:mccov6000} near here.]
         \end{center}
         \begin{center}
          [\cref{fig:tobinsqB1} near here.]
    \end{center}
    \begin{center}
          [\cref{fig:cashflowB1} near here.]
    \end{center}

       \newpage
        \begin{table}[ht]`
        \caption{}\vspace{-0.25cm}
        \centering
            \caption*{\textit{Simulation: bias and standard deviation ($n=6,000$).}\vspace{-0.5cm}}
            \label{table:mcn6000}
             \setlength{\tabcolsep}{2.6pt}
            \begin{center}
            \begin{tabular}{@{\extracolsep{4pt}}ll@{\hspace{\tabcolsep}}r@{\hspace{\tabcolsep}}rr@{\hspace{1\tabcolsep}}rr@{\hspace{1\tabcolsep}}rr@{\hspace{1\tabcolsep}}r}
               \hline
                &&\multicolumn{8}{c}{$\beta_{0}=0$, $\alpha_{0}=0.05$}\\
                \cline{3-10}
                &&\multicolumn{2}{c}{(1)}&\multicolumn{2}{c}{(2)}&\multicolumn{2}{c}{(3)}&\multicolumn{2}{c}{(4)}\\
                \cline{3-4}\cline{5-6}\cline{7-8}\cline{9-10}
OLS&$\widehat{\beta}$& 0.000 & (0.000) & 0.000 & (0.000) & 0.000 & (0.000) & 0.000 & (0.000) \\
&$\widehat{\alpha}$&0.050 & (0.001) & 0.050 & (0.001) & 0.050 & (0.001) & 0.050 & (0.001) \\[.5ex]
3M&$\widehat{\beta}$& 0.084 & (8.099) & -0.782 & (110.119) & 0.016 & (1.342) & 0.011 & (2.157) \\
&$\widehat{\alpha}$&-0.106 & (14.978) & 1.204 & (162.353) & 0.020 & (2.538) & 0.033 & (3.251) \\[.5ex]
DC(20)&$\widehat{\beta}$&0.001 & (0.008) & 0.000 & (0.009) & 0.001 & (0.007) & 0.001 & (0.008) \\
&$\widehat{\alpha}$&0.049 & (0.015) & 0.049 & (0.013) & 0.049 & (0.014) & 0.049 & (0.012) \\[.5ex]
DC(40)&$\widehat{\beta}$&0.001 & (0.005) & 0.001 & (0.005) & 0.001 & (0.005) & 0.001 & (0.005) \\
&$\widehat{\alpha}$&0.048 & (0.010) & 0.049 & (0.009) & 0.048 & (0.010) & 0.049 & (0.008) \\
                 \hline &&\multicolumn{8}{c}{$\beta_{0}=0.025$, $\alpha_{0}=0.05$}\\
                \cline{3-10} &&\multicolumn{2}{c}{(1)}&\multicolumn{2}{c}{(2)}&\multicolumn{2}{c}{(3)}&\multicolumn{2}{c}{(4)}\\
              \cline{3-4}\cline{5-6}\cline{7-8}\cline{9-10}
 OLS&$\widehat{\beta}$&0.011 & (0.000) & 0.009 & (0.000) & 0.011 & (0.000) & 0.009 & (0.000) \\
&$\widehat{\alpha}$&0.077 & (0.001) & 0.075 & (0.001) & 0.077 & (0.001) & 0.075 & (0.001) \\[.5ex]
3M&$\widehat{\beta}$&0.025 & (0.000) & 0.025 & (0.000) & 0.025 & (0.000) & 0.025 & (0.000) \\
&$\widehat{\alpha}$&0.050 & (0.001) & 0.050 & (0.001) & 0.050 & (0.001) & 0.050 & (0.001) \\[.5ex]
DC(20)&$\widehat{\beta}$&0.026 & (0.005) & 0.026 & (0.007) & 0.026 & (0.005) & 0.026 & (0.007) \\
&$\widehat{\alpha}$&0.049 & (0.010) & 0.049 & (0.011) & 0.049 & (0.010) & 0.049 & (0.011) \\[.5ex]
DC(40)&$\widehat{\beta}$&0.025 & (0.005) & 0.024 & (0.007) & 0.025 & (0.005) & 0.024 & (0.007) \\
&$\widehat{\alpha}$&0.049 & (0.010) & 0.052 & (0.010) & 0.049 & (0.009) & 0.051 & (0.010) \\
                \hline
                FE && \multicolumn{2}{c}{No} &  \multicolumn{2}{c}{Yes} &  \multicolumn{2}{c}{No} &  \multicolumn{2}{c}{Yes} \\
                TE && \multicolumn{2}{c}{No} &  \multicolumn{2}{c}{No} &  \multicolumn{2}{c}{Yes} &  \multicolumn{2}{c}{Yes} \\
               \hline
            \end{tabular}\vspace{-0.25cm}
            \end{center}
            \begin{minipage}{\textwidth}
       \textit{Note:} mean and standard deviation of the estimates for $\beta$ and $\alpha$ when the cross-sectional dimension is $n=6,000$. For details, see the note following \cref{table:mc}.
       \end{minipage}
        \end{table}
        \begin{table}
        \caption{}\vspace{-0.25cm}
        \centering
            \caption*{\textit{Simulation: coverage ($n=6,000$).}\vspace{-0.5cm}}
           \begin{center}
            \label{table:mccov6000}
            \begin{tabular}{@{\extracolsep{4pt}}llrrrrrrrr}
                \hline
                &&\multicolumn{4}{c}{$\beta_{0}=0$, $\alpha_{0}=0.05$}&\multicolumn{4}{c}{$\beta_{0}=0.025$, $\alpha_{0}=0.05$}\\
                \cline{3-6}\cline{7-10}
OLS &$\beta$& 95.1 & 94.3 & 95.0 & 94.3 & 0.0 & 0.0 & 0.0 & 0.0 \\
 &$\alpha$& 95.0 & 94.5 & 95.0 & 94.5 & 0.0 & 0.0 & 0.0 & 0.0 \\[.5ex]
3M &$\beta$& 97.0 & 98.4 & 97.0 & 98.4 & 95.0 & 95.5 & 95.0 & 95.5 \\
 &$\alpha$& 96.9 & 98.4 & 96.9 & 98.3 & 94.8 & 94.5 & 94.8 & 94.5 \\[.5ex]
DC(20) &$\beta$& 96.9 & 96.9 & 96.4 & 96.8 & 94.1 & 94.0 & 93.9 & 94.3 \\
 &$\alpha$ & 96.8 & 96.7 & 96.5 & 96.7 & 93.9 & 93.9 & 93.9 & 94.2 \\[.5ex]
DC(40) &$\beta$& 96.0 & 96.6 & 96.0 & 96.4 & 94.1 & 91.2 & 94.5 & 92.4 \\
 &$\alpha$& 95.9 & 96.3 & 95.8 & 96.2 & 93.9 & 91.0 & 94.4 & 92.2 \\
                \hline
                FE && \multicolumn{1}{c}{No} & \multicolumn{1}{c}{Yes} & \multicolumn{1}{c}{No} & \multicolumn{1}{c}{Yes}& \multicolumn{1}{c}{No} & \multicolumn{1}{c}{Yes} & \multicolumn{1}{c}{No} & \multicolumn{1}{c}{Yes}\\
                TE && \multicolumn{1}{c}{No} & \multicolumn{1}{c}{No} & \multicolumn{1}{c}{Yes} & \multicolumn{1}{c}{Yes}& \multicolumn{1}{c}{No} & \multicolumn{1}{c}{No} & \multicolumn{1}{c}{Yes} & \multicolumn{1}{c}{Yes}\\
                \hline
            \end{tabular}\vspace{-0.25cm}
            \end{center}
            \begin{minipage}{\textwidth}
       \textit{Note:} coverage rate ($\times 100$) for the model and methods as described below \cref{table:mc} with cross-sectional dimension $n=6,000$. For details, see the note following \cref{table:mccov}.
       \end{minipage}
        \end{table}

\clearpage

    \end{appendices}

\end{document}